# A Multi-timescale Two-stage Robust Grid-friendly Dispatch Model for Microgrid Operation

Jiayu Han[a], Lei Yan[a], Liuxi Zhang[b], Aleksi Paaso[b], Shay Bahramirad[b], Zuyi Li[a*]

[a] Illinois Institute of Technology, Electrical and Computer Engineering Department, Chicago, IL 60616 USA
[b] Commonwealth Edison Company, Oakbrook Terrace, IL 60181 USA
*Corresponding Author: Zuyi Li (e-mail: lizu@iit.edu).

## HIGHLIGHTS

- A multi-timescale two-stage robust grid-friendly dispatch model for microgrid operation is proposed.
- The model is tested for a community microgrid in a controlled hardware in loop testbed.
- The dispatch is robust as it can be immunized to both hourly solar and load uncertainties.
- The dispatch is grid-friendly as the combined solar-storage output can remain unchanged on an hourly basis.

## ARTICLE INFO

## ABSTRACT

Uncertainty in renewable energy generation and load consumption is a great challenge for microgrid operation, especially in islanded mode as the microgrid may be small in size and has limited flexible resources. In this paper, a multi-timescale, two-stage robust unit commitment and economic dispatch model is proposed to optimize the microgrid operation. The first stage is a combination of day-ahead hourly and real-time sub-hourly model, which means the day-ahead dispatch result must also satisfy the real-time condition at the same time. The second stage is to verify the feasibility of the day-ahead dispatch result in worst-case condition considering high-level uncertainty in renewable energy dispatch and load consumptions. In the proposed model, battery energy storage system (BESS) and solar PV units are integrated as a combined solar-storage system. The BESS plays an essential role to balance the variable output of solar PV units, which keeps the combined solar-storage system output unchanged on an hourly basis. In this way, it largely neutralizes the impact of solar uncertainty and makes the microgrid operation grid friendly. Furthermore, in order to enhance the flexibility and resilience of the microgrid, both BESS and thermal units provide regulating reserve to manage solar and load uncertainty. The model has been tested in a controlled hardware in loop (CHIL) environment for the Bronzeville Community Microgrid system in Chicago. The simulation results show that the proposed model works effectively in managing the uncertainty in solar PV and load and can provide a flexible dispatch in both grid-connected and islanded modes.

## Nomenclature

*Indices*

| | |
|---|---|
| $g$ | Thermal units $g = 1, \cdots, N_g$ |
| $b$ | BESS units $b = 1, \cdots, N_b$ |
| $dis$ | Discharging modes of BESS units |
| $c$ | Charging modes of BESS units |
| $pv$ | Solar PV units $pv = 1, \cdots, N_{pv}$ |
| $f$ | Forecasted value |
| $t$ | Hourly time slots $t = 1, \cdots, N_T$ |
| $\Delta$ | Sub-hourly time slots $\Delta = 1, \cdots, 60/dt$ |
| $ex$ | Exchanged power |
| $ld$ | Load shedding |

*Parameters*

| | |
|---|---|
| $P_g^{max}$ | Maximum power output of thermal unit $g$ |
| $P_g^{min}$ | Minimum power output of thermal unit $g$ |
| $RU_g$ | Maximum upward ramp of thermal unit $g$ |
| $RD_g$ | Maximum downward ramp of thermal unit $g$ |
| $T_{on,g}$ | Minimum ON time of thermal unit $g$ |
| $T_{off,g}$ | Minimum OFF time of thermal unit $g$ |
| $P_{pv,t}^f$ | Forecasted power output of solar PV unit $pv$ at time $t$ |
| $P_{pv,t,\Delta}^f$ | Forecasted power output of solar PV unit $pv$ during sub-hourly dispatch time slot $\Delta$ at time $t$ |
| $u_{pv,t}$ | Maximum uncertainty value of solar PV unit $pv$ at time $t$ |
| $\Delta$ | Sub-hourly time slot |
| $\delta$ | Sub-hourly time interval (5 minutes in this proposed model) |
| $N_b$ | Number of BESS units |
| $N_{pv}$ | Number of solar PV units |
| $N_T$ | Number of time slots in the whole dispatchable period |
| $N_{seg}$ | Number of sub-hourly time slots in every hour |
| $\eta_{b,c}$ | Charging efficiency of BESS unit $b$ |
| $\eta_{b,dis}$ | Discharging efficiency of BESS unit $b$ |
| $P_{b,max}^{dis}$ | Maximum discharging power output of BESS unit $b$ |
| $P_{b,max}^c$ | Maximum charging power output of BESS unit $b$ |
| $\alpha$ | Solar PV output uncertainty rate, $0 \leq \alpha \leq 1$ |
| $\beta$ | Load uncertainty rate, $0 \leq \beta \leq 1$ |

*Variables*

| | |
|---|---|
| $C_g^P(\cdot)$ | Cost function of dispatched power of thermal unit $g$ |
| $C_g^I(\cdot)$ | Start-up and shutdown cost function of thermal unit $g$ |
| $C_g^r(\cdot)$ | Cost function of regulating reserve of thermal unit $g$ |
| $C_b^P(\cdot)$ | Cost functions of charging and discharging power of BESS unit $b$ |
| $C_b^r(\cdot)$ | Cost functions of regulating reserve of BESS unit $b$ |
| $C_{ex}^P(\cdot)$ | Cost function of power exchange between the utility grid and microgrid |
| $C_{LC}^P(\cdot)$ | Cost function of load curtailment |
| $P_{g,t}$ | Power output of thermal unit $g$ at time $t$ |
| $ru_{g,t}$ | Upward regulating reserve of thermal unit $g$ at time $t$ |
| $rd_{g,t}$ | Downward regulating reserve of thermal unit $g$ at time $t$ |
| $I_{g,t}$ | On/off status of thermal unit $g$ at time $t$ |
| $y_{g,t}$ | Startup indicator of thermal unit $g$ at time $t$ |
| $z_{g,t}$ | Shutdown indicator of thermal unit $g$ at time $t$ |
| $P_{pv,t}$ | Power output of solar PV unit $pv$ at time $t$ |
| $P_{pv,t,\Delta}$ | Power output of solar PV unit $pv$ during sub-hourly dispatch time slot $\Delta$ at time $t$ |
| $P_{b,t}^{dis}$ | Discharging power output of BESS unit $b$ at time $t$ |
| $P_{b,t}^c$ | Charging power output of BESS unit $b$ at time $t$ |
| $LC_t$ | The value of load curtailment in islanded mode at time $t$ |
| $\widehat{LC}_{t,\Delta}$ | The value of load curtailment during sub-hourly dispatch time slot $\Delta$ at time $t$ |
| $ru_{b,t}^{dis}$ | Upward regulating reserve of BESS unit $b$ on discharging mode at time $t$ |
| $rd_{b,t}^{dis}$ | Downward regulating reserve of BESS unit *b* on discharging mode at time *t* |
| $ru_{b,t}^{c}$ | Upward regulating reserve of BESS unit *b* on charging mode at time *t* |
| $rd_{b,t}^{c}$ | Downward regulating reserve of BESS unit *b* on charging mode at time *t* |
| $I_{b,t}^{dis}$ | Discharging status of BESS unit *b* at time *t* |
| $I_{b,t}^{c}$ | Charging status of BESS unit *b* at time *t* |
| $E_{b,t}$ | Stored energy of BESS unit *b* at time *t* |
| $E_{b,t,\Delta}$ | Stored energy of BESS unit *b* during sub-hourly dispatch time slot *Δ* at time *t* |
| $\varepsilon_{pv,t}$ | Uncertainty value of solar PV unit *pv* at time *t* |
| $\varepsilon_{ld,t}$ | Uncertainty value of load at time *t* |

## 1. Introduction

Sustainability, security and environmental protection are three important goals for modern power systems. To achieve those goals, more and more renewable energy sources (RES) are used and small-scale localized power systems like microgrids are being investigated in recent years. A microgrid consists of conventional and dispatchable generation sources, RES, battery energy storage system (BESS) and loads, which can be coordinated to operate in both grid-connected and islanded modes.

Uncertainty in renewable energy generation and load consumption is always a big issue for the secure operation of any power system, be it large or small. Uncertainty is especially a great challenge for microgrid operation as the size of a microgrid is generally small and has limited flexible resources. To solve the uncertainty problem in microgrid, energy storages are largely used. Other than dealing with uncertainties, energy storages also introduce many other advantages like balancing generation and demand, power quality improvement, smoothing the renewable resource's intermittency, and enabling ancillary services like frequency and voltage regulation in microgrid [1]. BESS, as a key part of energy storages, has been used to manage the uncertain power output from solar PV in microgrid. There are plenty of researches about BESS in microgrid planning and operation. Ref. [2] utilizes BESS in addition to wind-PV generation system in microgrid to increase operation reliability and economic benefits. Two algorithms are used in [3] to optimize the size of wind turbine, solar PV panels, and BESS in the grid-connected microgrid. Ref. [4] uses a dual battery bank to manage the power balance and voltage of the isolated microgrid; the first bank helps meet the load demand and the second one is used to regulate voltage. Ref. [5] presents a power dispatch model for BESS that can be used in both grid-connected and islanded modes. Ref. [6] uses semidefinite programming to optimize the economic dispatch of a dc microgrid with high penetration of distributed generations and energy storage systems. However, in the above references, BESS is mainly used to provide energy and power to balance the power of the whole microgrid but not used to manage or balance the output power of a certain type of RES. In this paper, a robust dispatch model is proposed for the microgrid operation and can help mitigate the uncertainty issue of RES (e.g. solar PV units). One of the major innovations of the proposed model is the utilization of solar PV and energy storage systems in a combined fashion (i.e., solar-storage system). There have been few reported works on the optimal integration and the utilization of such systems in an islanded microgrid by coordinating their operation with other microgrid elements. The BESS unit in the solar-storage system is prepared to provide energy to balance the uncertain output of solar PV unit under any scenario (e.g. in both grid-connected and islanded modes). Thus, the solar-storage system maintains a constant output on an hourly basis that is equal to the forecasted hourly output of solar PV units, and reduces the undesired impact on the utility grid caused by the uncertainty output of solar PV units.

Two common methods to address the uncertainty problem are scenario-based stochastic optimization and robust optimization [7]-[17]. Stochastic optimization [10]-[14] generates a large number of scenarios as a proxy to model the uncertainty. However, the drawback of this method is that it is computationally intensive in general and its efficacy needs to be improved [10]. Another useful method of dealing with uncertainty is robust optimization. Most works on robust optimization of power system applications are based on a two-stage optimization model [7]-[9]. A non-conservative day-ahead robust security-constrained unit commitment (SCUC) model with adjustment cost (or recourse cost) constraint is presented in [7]-[8]. Ref. [9] uses day-ahead robust SCUC model to accommodate wind output uncertainty. Most robust SCUC is focused on a single-timescale model, i.e. either day-ahead or real-time model. However, since RES has large uncertainty and more complex conditions than what can be considered in day-ahead dispatch, a single-timescale model is less flexible and less reliable than a multi-timescale model. There have been some studies on multi-timescale models [18]-[26]. Two types of multi-timescale models are considered in recent researches. The first type of multi-timescale models uses a two-stage optimization model, where the first stage is an hourly dispatch and the second stage is a sub-hourly dispatch [18]-[21]. Ref. [18] proposes a coordinated multi-timescale robust scheduling framework for isolated system with energy storage units. A multi-timescale rolling optimal dispatching framework is used in [19]. Ref. [20] proposes an optimal coordination strategy, which uses a multi-timescale model for isolated power systems. Ref. [21] discusses home energy management (HEM) with a multi-timescale optimization model. The second type of the multi-timescale model uses two optimization models: the first model is a day-ahead unit commitment model and the second one is a real-time dispatch model [22]-[23] to achieve the goal of multi-time scale. Ref. [22] proposes a multi-timescale and robust energy management and the scheme is divided into a day-ahead and an intraday model. The day-ahead model determines the baseline transaction between utility grid and microgrid and the intraday model determines the economic dispatch. The day-ahead scheduling model in [23] considers uncertainty of wind and solar power by multi-scenarios and applies dispatch schemes of different timescales in real-time dispatch. The multi-timescale model proposed in this paper is different from the two types mentioned above, although the proposed multi-timescale model also uses a two-stage robust model. The one feature of the proposed model, which is very different from Ref. [10]-[21], is that the first stage of the proposed model considers not only hourly dispatch but also sub-hourly dispatch, and the second stage considers the worst case of hourly dispatch. Therefore, the day-ahead dispatch results can not only satisfy any day-ahead conditions but also can deal with some real-time situations. Compared to Ref. [22]-[23], the proposed model can get hourly unit commitment and sub-hourly dispatch within one model, which is time saving and more efficient.

Nowadays, artificial intelligence-based methods have been under rapid development and they are also introduced to solve the problems in power systems. Most machine learning methods including deep learning methods are used to forecast load, renewable energy, and electricity price because they can discover the inherent nonlinear features and high-level invariant structures in data [27]. Ref. [28] uses long short-term memory neural networks to forecast future load and wind power. Ref. [29] addresses the issue of ultra-short-term wind power time series forecasting based on extreme learning machine. Furthermore, reinforcement learning is used for energy management with uncertainties. Ref. [30] provides operation strategies considering uncertainties based on double deep Q-learning method. However, machine learning methods need a large number of data to train a model and these models are easy to be attacked, so the models are less robust and the results are less reliable [31]. In comparison, the model proposed in this paper does not need so many data to train. In addition, the model-based algorithm itself (e.g. robust optimization) is less affected by the quality of the data so it is not easy to be attacked. Therefore, the proposed model is more time-saving and safer.

Furthermore, hourly regulating reserve is considered in this model, which increases the flexibility and security of the microgrid system. Therefore, the results can be used not only for day-ahead dispatch considering uncertainty in both grid-connected and islanded modes but also prepare enough regulating reserve to satisfy real-time dispatch constraints.

There are four major features of the proposed model:
- **Multi-timescale model**: This is important since the solar PV unit output may change rapidly within one hour. In comparison, most traditional models are single-timescale models (day-ahead hourly model or real-time model). The proposed model is suitable for real-time applications.
- **Robust model**: In general, solar PV unit output cannot be forecasted very accurately, and the forecasting error could be fairly high (e.g., 15%). The robust model can deal with the uncertainty of both PV unit and load. In comparison, most traditional models do not explicitly model the uncertainty.
- **Constant hourly aggregated solar-storage output**: The aggregated output of the solar-storage system in the proposed model will remain unchanged on an hourly basis even if the actual solar PV unit output deviates from the forecasted values. This is important if (1) there is no thermal unit in a microgrid; or (2) the thermal units in a microgrid are not fast enough to track the rapid change of solar PV unit output; or (3) the microgrid is in the islanded mode and does not have utility grid support.
- **Regulating reserve**: The regulating reserve is used to deal with uncertainty, which increases the flexibility and resilience of the microgrid. In this paper, both BESS units and thermal units can provide regulating reserve. The regulating reserve of BESS unit will manage the uncertainty of solar PV unit and the regulating reserve of thermal units will manage the uncertainty caused by loads.

The rest of this paper is organized as follows. Section 2 describes the mathematical formulation and solution approach of the proposed multi-timescale robust unit commitment (RUC) model. Section 3 presents case studies to examine the effectiveness of the proposed model. Section 4 concludes this paper.

**2. Mathematical Formulation and Solution Approach**

This section presents the mathematical formulation of the proposed model and corresponding solution approach.

*2.1 Multi-timescale two-stage RUC model*

The traditional robust unit commitment (RUC) model is a single-timescale model, which only considers hourly dispatch or sub-hourly dispatch. In this paper, a multi-timescale two-stage RUC model is proposed, which integrates the day-ahead RUC model and sub-hourly economic model to satisfy the continuously changing output power of solar PV units and load demand. This is important since the PV unit output may change rapidly within one hour. Therefore, the results of the proposed model not only provide enough power and energy for supplying the forecasted load demand in both hourly and sub-hourly dispatches but also has sufficient amount of reserve to deal with the worst-case uncertain condition of load and solar PV unit.

The proposed multi-timescale two-stage RUC model includes three parts: the hourly RUC model, the sub-hourly dispatch model, and the hourly worst-case dispatch model. The uncertainty of solar PV units and load that motivates the proposed model is discussed first, followed by the objective function and the three individual models.

*2.1.1 Uncertainty modeling*

The output of solar PV units can change within one hour due to varying weather conditions and load demand may also have fluctuations due to various environmental and socio-economic factors. Therefore, two kinds of uncertainty sets are considered in this proposed model. The first set is the uncertainty set of load demand and the second one is that of solar PV units. It is assumed that the time scales of the uncertainty set of the load demands and the solar PV units are both one hour.

The uncertainty sets of solar PV units and load demand are represented as follows:

$$U_{pv} := \{\varepsilon_{pv} \in \mathbb{R}^{N_{pv}N_T}: -u_{pv,t} \leq \varepsilon_{pv,t} \leq u_{pv,t}, \forall pv, t\} \quad (1)$$

$$U_{Load} := \{\varepsilon_{ld} \in \mathbb{R}^{N_T}: -u_{ld,t} \leq \varepsilon_{ld,t} \leq u_{ld,t}, \forall ld, t\} \quad (2)$$

where, $u_{pv,t} = \alpha \cdot P_{pv,t}$ and $u_{ld,t} = \beta \cdot Load_t$. $0 \leq \alpha \leq 1$ and $0 \leq \beta \leq 1$.

As uncertainty widely occurs in a microgrid system due to constantly changing weather conditions, more reserves are needed to improve the microgrid reliability and flexibility. Regulating reserve can provide instantaneous power to balance power system supply and demand. In the proposed model, regulating reserve of thermal units is used to manage the load uncertainty in hourly dispatch while BESS units provide regulating reserves to deal with the uncertainty of solar PV units. Furthermore, the BESS and solar PV system are integrated as a solar-storage system. The aggregated output of the solar-storage system remains unchanged even if the outputs of solar PV units deviate from the forecasted values, which is grid-friendly from the utility grid's perspective. Thus, the BESS units should have sufficient regulating reserve available to achieve that goal.

### 2.1.2 Objective function

$$\min C_T = \begin{cases} \sum_t \begin{pmatrix} \sum_g C_g^I(P_{g,t}) + C_{ex}^P(P_{ex,t}) + C_{LC}^P(LC_t) \\ + \sum_g C_g^r(ru_{g,t} + rd_{g,t}) \\ + \sum_b C_b^r(ru_{b,t}^{dis} + ru_{b,t}^c) \end{pmatrix} \\ + \sum_t \sum_\Delta \begin{pmatrix} \sum_g C_g^P(\hat{P}_{g,t,\Delta}) \cdot \delta \\ + \sum_b C_b^P(\hat{P}_{b,t,\Delta}^{dis} + \hat{P}_{b,t,\Delta}^c) \cdot \delta \\ + C_{LC}^P(s_{LC,t,\Delta}) \cdot \delta \end{pmatrix} \\ + \sum_t Z_t \end{cases} \quad (3)$$

The objective function of the proposed model is shown in (3). The first three lines are hourly commitment/dispatch cost, which includes startup/shutdown cost of thermal units, exchanged power cost in grid-connected mode, load curtailment cost in islanded mode (Line 1), regulating reserve cost of thermal units (Line 2) and regulating reserve cost of BESS units (Line 3). The thermal units may not necessarily participate in optimizing microgrid operation for the purpose of cost minimization due to regulatory reasons. The next three lines are sub-hourly dispatch cost, which includes sub-hourly dispatch cost of thermal units (Line 4), sub-hourly dispatch cost of BESS units (Line 5), and penalty load curtailment cost in sub-hourly dispatch (Line 6). The last line is the recourse cost due to load curtailment in worst-case dispatch (Line 7).

### 2.1.3 Hourly RUC model

The hourly RUC model is shown in (4)-(30).

$$P_g^{min} I_{g,t} \leq P_{g,t} \leq P_g^{max} I_{g,t}, \forall g, t \quad (4)$$
$$P_{g,t} - P_{g,t-1} \leq RU_g(1 - y_{g,t}) + P_g^{min} y_{g,t}, \forall g, t \quad (5)$$
$$P_{g,t-1} - P_{g,t} \leq RD_g(1 - z_{g,t}) + P_g^{min} z_{g,t}, \forall g, t \quad (6)$$
$$\sum_{\tau=t}^{t+T_{on,g}-1} I_{g,\tau} \geq T_{on,g} y_{g,t}, \forall g, t \quad (7)$$
$$\sum_{\tau=t}^{t+T_{off,g}-1} I_{g,\tau} \leq T_{off,g}(1 - z_{g,t}), \forall g, t \quad (8)$$
$$y_{g,t} + z_{g,t} \leq 1, \forall g, t \quad (9)$$
$$y_{g,t} - z_{g,t} = I_{g,t} - I_{g,t-1}, \forall g, t \quad (10)$$
$$P_{g,t} + ru_{g,t} \leq P_g^{max} I_{g,t}, \forall g, t \quad (11)$$
$$P_{g,t} - rd_{g,t} \geq P_g^{min} I_{g,t}, \forall g, t \quad (12)$$
$$0 \leq ru_{g,t} \leq RU_g I_{g,t}, \forall g, t \quad (13)$$
$$0 \leq rd_{g,t} \leq RD_g I_{g,t}, \forall g, t \quad (14)$$
$$0 \leq P_{pv,t} \leq P_{pv,t}^f, \forall pv, t \quad (15)$$
$$-P_{ex}^{max} I_{ex} \leq P_{ex,t} \leq P_{ex}^{max} I_{ex}, \forall t \quad (16)$$
$$-k_{ex} P_{ex}^{max} \leq P_{ex,t} - P_{ex,t-1} \leq k_{ex} P_{ex}^{max}, 0 \leq k_{ex} \leq 1, \forall t \quad (17)$$
$$0 \leq LC_t \leq (Load_t - Ld_{VIP,t})(1 - I_{ex}), \forall t \quad (18)$$
$$0 \leq P_{b,t}^{dis} \leq P_{b,max}^{dis} I_{b,t}^{dis}, \forall b, t \quad (19)$$
$$0 \leq P_{b,t}^c \leq P_{b,max}^c I_{b,t}^c, \forall b, t \quad (20)$$
$$I_{b,t}^{dis} + I_{b,t}^c \leq 1, \forall b, t \quad (21)$$
$$P_{b,t}^{dis} + ru_{b,t}^{dis} \leq P_{b,max}^{dis} I_{b,t}^{dis}, \forall b, t \quad (22)$$
$$P_{b,t}^c + ru_{b,t}^c \leq P_{b,max}^c I_{b,t}^c, \forall b, t \quad (23)$$
$$ru_{b,t}^{dis} \geq 0, ru_{b,t}^c \geq 0, \forall b, t \quad (24)$$
$$E_{b,t} = E_{b,t-1} + \eta_{b,c} P_{b,t}^c - P_t^{dis}/\eta_{b,dis}, \forall b, t \quad (25)$$
$$E_b^{min} \leq E_{b,t} \leq E_b^{max}, \forall b, t \quad (26)$$
$$E_{b,t} - ru_{b,t}^{dis}/\eta_{b,dis} \geq E_b^{min}, \forall b, t \quad (27)$$
$$E_{b,t} + \eta_{b,c} ru_{b,t}^c \leq E_b^{max}, \forall b, t \quad (28)$$
$$E_{b,T} = E_{b,0}, \forall b \quad (29)$$
$$\sum_g P_{g,t} + \sum_{pv} P_{pv,t} + \sum_b (P_{pv,t}^{dis} - P_{pv,t}^c) + P_{ex,t} + LC_t = Load_t, \forall t \quad (30)$$

Constraints (4)-(14) are for thermal units. The minimum and maximum output powers of thermal units are limited in (4). Ramping up/down constraints are shown in (5)-(6). (7)-(8) are the minimum ON/OFF time constraints. The relationships of $y_{g,t}$, $z_{g,t}$, and $I_{g,t}$ are modeled in (9)-(10). The feasible output power range including regulating reserve is shown in (11)-(12). The regulating reserve should not exceed the ramping capability of the thermal units and should not be negative as presented in (13)-(14).

Eq. (15) is the output power range of solar PV units. In the proposed model, a solar PV unit can be dispatchable if required and its maximum output is the forecasted value. Constraints (16)-(17) are the range of the power exchange between the utility grid and the microgrid. To avoid large fluctuations and have a grid-friendly dispatch, the hourly change of exchanged power should be limited, which is the first aspect of the "grid-friendly" dispatch. Eq. (18) is the range of load shedding in the islanded mode. $Ld_{VIP,t}$ is the critical load that cannot be curtailed in the islanded mode.

Constraints (19)-(29) are for BESS units. The range of discharging and charging power of BESS units is presented in (19)-(20). BESS units can only be in one mode at any given time, either discharging mode or charging mode as shown in (21). In this model, BESS units can provide upward regulating reserve in both discharging and charging modes. Feasible output range of discharging and charging power including regulating reserve is shown in (22)-(23) and regulating reserves should not be negative as shown in (24). BESS energy constraints are presented in (25)-(29). The total energy stored in a BESS unit at time $t$ is presented by (25) and it cannot exceed the energy range as shown in (26). It is stated in (27)-(28) that BESS units need to have enough energy to provide regulating reserve. It is assumed in (29) that the energy at the end of the scheduling horizon ($t = T$) in the BESS units must be the same as that in the beginning of the scheduling horizon ($t = 0$). Power balance constraint in the day-ahead dispatch of the microgrid is shown in (30).

### 2.1.4 Sub-hourly dispatch model

The sub-hourly dispatch model is shown in (31)-(44).

$$P_{g,min} I_{g,t} \leq \hat{P}_{g,t,\Delta} \leq P_{g,max} I_{g,t}, \forall g, t, \Delta \quad (31)$$
$$-rd_{g,t} \leq \hat{P}_{g,t,\Delta} - P_{g,t} \leq ru_{g,t}, \forall g, t, \Delta \quad (32)$$
$$\hat{P}_{pv,t,\Delta} = P_{pv,t,\Delta}^f, \forall pv, t, \Delta \quad (33)$$
$$0 \leq \hat{P}_{b,t,\Delta}^{dis} \leq P_{b,max}^{dis} I_{b,t,\Delta}^{dis}, \forall b, t, \Delta \quad (34)$$
$$0 \leq \hat{P}_{b,t,\Delta}^c \leq P_{b,max}^c I_{b,t,\Delta}^c, \forall b, t, \Delta \quad (35)$$
$$I_{b,t,\Delta}^{dis} + I_{b,t,\Delta}^c \leq 1, \forall b, t, \Delta \quad (36)$$
$$\hat{E}_{b,t,\Delta} = \hat{E}_{b,t,\Delta-1} + \eta_c \cdot \delta \cdot \hat{P}_{b,t,\Delta}^c - 1/\eta_{dis} \cdot \delta \cdot \hat{P}_{b,t,\Delta}^{dis}, \forall b, t, \Delta \quad (37)$$
$$E_b^{min} \leq \hat{E}_{b,t,\Delta} \leq E_b^{max}, \forall b, t, \Delta \quad (38)$$
$$-P_{ex}^{max} I_{ex} \leq \hat{P}_{ex,t,\Delta} \leq P_{ex}^{max} I_{ex}, \forall t, \Delta \quad (39)$$
$$0 \leq \widehat{LC}_{t,\Delta} \leq (Load_{t,\Delta} - Ld_{VIP,t,\Delta}) \cdot (1 - I_{ex}), \forall t, \Delta \quad (40)$$
$$0 \leq \widehat{LC}_{t,\Delta} \leq LC_t + s_{LC,t,\Delta}, \forall t, \Delta \quad (41)$$
$$s_{LC,t,\Delta} \geq 0, \forall t, \Delta \quad (42)$$
$$\sum_{pv} \hat{P}_{pv,t,\Delta} + \sum_b (\hat{P}_{b,t,\Delta}^{dis} - \hat{P}_{b,t,\Delta}^c) = \sum_{pv} P_{pv,t}, \forall t, \Delta \quad (43)$$
$$\sum_g \hat{P}_{g,t,\Delta} + \sum_{pv} \hat{P}_{pv,t,\Delta} + \sum_b (\hat{P}_{b,t,\Delta}^{dis} - \hat{P}_{b,t,\Delta}^c) + \hat{P}_{ex,t,\Delta} + \widehat{LC}_{t,\Delta} = Load_{t,\Delta}, \forall t, \Delta \quad (44)$$

where $\delta$ is the time interval of sub-hourly dispatch and in this model $\delta = 5/60$ h. The output power of a thermal unit needs to be within its capacity limits as shown in (31). The output power of the thermal units in sub-hourly model might be different from that in the base case (day-ahead dispatch) due to load uncertainty. This leads to the difference between the load in day-ahead forecast and the load in sub-hourly forecast, but the change should be less than the regulating reserve as presented in (32). The output of a solar PV unit in sub-hourly dispatch model is equal to the sub-hourly forecasted value shown in (33). The range of discharging and charging power of BESS units in the sub-hourly dispatch are represented in (34) and (35). As shown in (36), the operating status of BESS units in sub-hourly dispatch can be different from that in day-ahead dispatch. The energy constraints of the BESS are presented in (37)-(38). The energy stored in the BESS units in sub-hourly dispatch is shown in (37) and its range is shown in (38). The constraints of the exchanged power and load curtailment are shown in (39)-(41). The output values of exchanged power should be within its range as shown in (39). The range of load shedding in the islanded mode is shown in (40). Variable $s_{LC,t,\Delta}$ in (41) and (42) is a slack variable used to relax the limit on sub-hourly load curtailment. In this way, on one hand, the operation of load curtailment is more flexible in sub-hourly dispatch. On the other hand, it can still guarantee the overall economical operation as a penalty cost would incur if additional load curtailment is needed. The total output of the solar-storage system in sub-hourly dispatch must be equal to the total output of solar PV units in hourly dispatch when solar PV unit output is nonzero (e.g., from 7am to 6pm) as shown in (43), which is the second aspect of the "grid-friendly" dispatch. The total power balance in sub-hourly dispatch is represented in (44).

### 2.1.5 Worst-case dispatch model

The worst-case dispatch model is shown in (45)-(60).

$$P_{g,min}I_{g,t} \leq \hat{P}_{g,t}^w \leq P_{g,max}I_{g,t}, \forall g, t \tag{45}$$
$$-rd_{g,t} \leq \hat{P}_{g,t}^w - P_{g,t} \leq ru_{g,t}, \forall g, t \tag{46}$$
$$\hat{P}_{pv,t}^w = P_{pv,t}^f + \varepsilon_{pv,t}, \forall pv, t \tag{47}$$
$$0 \leq \hat{P}_{b,t}^{w,dis} \leq P_{b,max}^{dis} I_{b,t}^{dis}, \forall b, t \tag{48}$$
$$0 \leq \hat{P}_{b,t}^{w,c} \leq P_{b,max}^{c} I_{b,t}^{c}, \forall b, t \tag{49}$$
$$\hat{P}_{b,t}^{w,dis} \leq P_{b,t}^{dis} + ru_{b,t}^{dis}, \forall b, t \tag{50}$$
$$\hat{P}_{b,t}^{w,c} \leq P_{b,t}^{c} + ru_{b,t}^{c}, \forall b, t \tag{51}$$
$$\hat{E}_{b,t}^w = \hat{E}_{b,t-1}^w + \eta_c \cdot \hat{P}_{b,t}^{w,c} - 1/\eta_{dis} \cdot \hat{P}_{b,t}^{w,dis}, \forall b, t \tag{52}$$
$$\hat{E}_{b,T}^w = E_{b,0}, \forall b, t \tag{53}$$
$$E_b^{min} \leq \hat{E}_{b,t}^w \leq E_b^{max}, \forall b, t \tag{54}$$
$$-P_{ex}^{max}I_{ex} \leq \hat{P}_{ex,t}^w \leq P_{ex}^{max}I_{ex}, \forall t \tag{55}$$
$$0 \leq \widehat{LC}_t^w \leq (Load_t - Ld_{VIP,t}) \cdot (1 - I_{ex}), \forall t \tag{56}$$
$$C_{LC}^P(\widehat{LC}_t^w) \leq Z_t, \forall t \tag{57}$$
$$Z_t \geq 0, \forall t \tag{58}$$
$$\sum_g \hat{P}_{g,t}^w + \sum_{pv} \hat{P}_{pv,t}^w + \sum_b (\hat{P}_{b,t}^{w,dis} - \hat{P}_{b,t}^{w,c})$$
$$+ \hat{P}_{ex,t}^w + \widehat{LC}_t^w = Load_t + \varepsilon_{ld,t}, \forall t \tag{59}$$
$$\sum_{pv} \hat{P}_{pv,t}^w + \sum_b (\hat{P}_{b,t}^{w,dis} - \hat{P}_{b,t}^{w,c}) = \sum_{pv} P_{pv,t}, \forall t \tag{60}$$

The output power of the thermal units is limited in (45)-(46). Output uncertainty of solar PV units and that of load demand are shown in (47) and (59), respectively. The output power of BESS units in the worst-case condition is limited in (48)-(49), and also should not exceed the sum of the output in day-ahead dispatch and regulating reserve (50)-(51). In the proposed model, only short term (e.g. day-ahead hourly and sub-hourly) unit commitment and dispatch is considered and the degradation of battery units is ignored in this paper. Thus, the maximum value of discharging/charging power is assumed to be fixed. The energy constraints of the BESS units are shown in (52)-(54). The limits of exchanged power with the utility grid and load curtailment are shown in (55)-(56). The recourse cost constraint of load curtailment in worst case is shown in (57) and it should be non-negative as shown in (58). The total power balance is represented in (59). In (60), the combined output value of the solar-storage system needs to remain the same as the forecasted value of hourly PV output for hours when solar PV unit output may be nonzero, which ensures the "grid-friendly" dispatch.

### 2.2 Multi-timescale two-stage RUC solution

The compact form of the proposed multi-timescale RUC model (3)-(60) is:

$$\min_{(P,x)\in\phi} C(P,\hat{P},x) \tag{61}$$
$$\text{s.t.} \quad A \cdot P + B \cdot x + C \cdot \hat{P} \leq b \tag{62}$$

and

$$\Phi := \{(P,x): \forall \varepsilon \in U, \exists \hat{P}^w, \text{such that}$$
$$D_1 \cdot \hat{P}^w + D_2 \cdot P + E \cdot \varepsilon \leq h \tag{63}$$
$$F \cdot \hat{P}^w + G \cdot P + H \cdot x \leq g\} \tag{64}$$

where (61) represents (3); binary variable vector $x$ represents commitment variables of base-case condition; $P$ denotes dispatch variables of day-ahead dispatch; $\hat{P}$ denotes dispatch variables of sub-hourly dispatch; $\hat{P}^w$ stands for the adjusted generation in worst-case condition due to uncertainty; $A$, $B$ and $C$ are abstract matrices representing constraints (4)-(30) and (31)-(44). $D_1$, $D_2$, $E$, $F$, $G$ and $H$ are abstract matrices representing constraints (45)-(60).

By applying Column Generation (CG) method, the multi-timescale RUC is divided into two parts: Master Problem (MP) and Subproblem (SP), which is shown as follows:

$$\text{(MP)} \quad \min_{(P,\hat{P},x)} C(P,\hat{P},x) \tag{65}$$
$$\text{s.t.} \quad A \cdot P + B \cdot x + C \cdot \hat{P} \leq b \tag{66}$$
$$D_1 \cdot \hat{P}^{w,v} + D_2 \cdot P \leq h - E \cdot \varepsilon^v, \forall v \in V \tag{67}$$
$$F \cdot \hat{P}^{w,v} + G \cdot P + H \cdot x \leq g \tag{68}$$

and

$$\text{(SP)} \quad \Omega := \max_{\varepsilon \in U} \min_{(s,\hat{P}^w)} 1^T s \tag{69}$$
$$\mathcal{F}(\varepsilon) = \{(s,\hat{P}^w): s \geq 0$$
$$D_1 \cdot \hat{P}^w - s \leq h - E \cdot \varepsilon - D_2 \cdot P \tag{70}$$
$$F \cdot d\hat{P}^v \leq g - G \cdot P - H \cdot x\} \tag{71}$$

where $V$ is the index set for uncertainty points which are dynamically generated in SP during the iterative solution process. Duality theory is used to convert the max-min problem to a maximization problem. The converted problem is shown as follows:

$$\text{(BP)} \quad \Omega = \max_{\varepsilon \in U, \lambda, \mu} \bar{h}_{dp}^T \lambda - (E\varepsilon)^T \lambda + \bar{g}_{fk}^T \mu \tag{72}$$
$$\text{s.t.} \quad D_1^T \cdot \lambda + F^T \cdot \mu = 0 \tag{73}$$
$$-1 \leq \lambda \leq 0, \mu \leq 0 \tag{74}$$

where, $\bar{h}_{dp} = h - D_2 \cdot P$ and $\bar{g}_{fk} = g - G \cdot P - H \cdot x$. The method to solve this problem is the same as that in [5].

## 3. Case Study

The proposed multi-timescale RUC model is tested in the Bronzeville Community Microgrid (BCM), which is being built in Chicago's Bronzeville community that can increase reliability, save costs and reduce carbon footprint. The BCM is designed to be equipped with the proposed model, which aims to be an enabling technology for the widespread sustainable deployment of low-cost, flexible, and reliable PV generation. The BCM setting provides an excellent opportunity to leverage and demonstrate the merits of the solar PV/battery storage investment to achieve better economic, resilience, and reliability benefits of solar PV. The real-time digital power system simulator (RTDS) is used to simulate the real-time BCM operation and to verify the effectiveness of the proposed model while the construction of BCM is in progress. The simplified one-line diagram for the BCM is shown in Fig. 1.

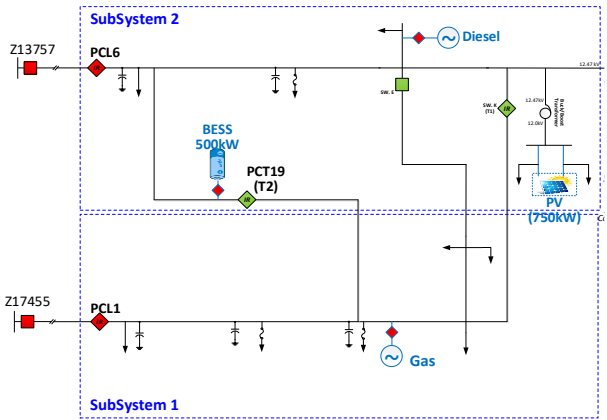

**Fig. 1**. Simplified one-line diagram for BCM

The peak load of the BCM is about 7 MW. In the simulated model, there are two thermal units (one gas unit and one diesel unit), one BESS unit, and one solar PV unit in the BCM. The capacities of the gas unit and the diesel unit are 6 MW and 3 MW, respectively. The capacities of the solar PV unit and the BESS unit are 750 kW and 500kW/2000 kWh, respectively. Other unit parameters are shown in Table 1, Table 2 and Table 3. The forecasted day-ahead and real-time values of the solar PV unit output and the load in the case study are shown in Fig. 2. The MIP problem is solved using Gurobi 7.5.2 on PC with Inter i7-6700 CPU @ 3.4 GHz 8GB RAM.

### 3.1 Simulation results in grid-connected mode

In this section, we examine the effectiveness of the proposed model for BCM in the grid-connected mode. There are two types of uncertainty sets in the proposed model: solar PV and load within a scale of one hour. $\alpha$ and $\beta$ are the uncertainty rates of the solar PV and the load, respectively. The goal is to provide day-ahead unit commitment and power dispatch that can satisfy any uncertain condition.

**Table 1**
Parameters of the gas unit and the diesel unit

| Unit | a ($) | b ($/kWh) | c ($/kWh²) | Pmin (kW) | Pmax (kW) | SU ($) | SD ($) |
|---|---|---|---|---|---|---|---|
| G1 | 50 | 0.07 | 5*10⁻⁶ | 200 | 6000 | 100 | 20 |
| G2 | 60 | 0.20 | 6*10⁻⁶ | 100 | 3000 | 300 | 20 |

| Unit | Min on(h) | Min off(h) | RU (kW) | RD (kW) | I0 | IH (h) | P0 (kW) |
|---|---|---|---|---|---|---|---|
| G1 | 4 | 4 | 2000 | 2000 | 1 | 4 | 800 |
| G2 | 3 | 2 | 1500 | 1500 | 1 | 3 | 800 |

a,b,c: coefficients of quadratic cost function Cost=a+b*P+c*P*P; SU: startup cost; SD: shutdown cost; RU: ramp-up rate; RD: ramp-down rate; I0: initial status; IH: initial hour; P0: initial dispatch

**Table 2**
Parameters of the BESS unit

| Emin (kWh) | Emax (kWh) | Ecap (kWh) | E0 (kWh) |
|---|---|---|---|
| 200 | 1800 | 2000 | 500 |
| Pc,max (kW) | Pdis,max (kW) | Charging efficiency | Discharging efficiency |
| 500 | 500 | 0.95 | 0.95 |

**Table 3**
Values of prices

| Price of exchanged power ($/kWh) | Price of charging power ($/kWh) | Price of discharging power ($/kWh) | Price of load curtailment ($/kWh) | Price of regulating reserve ($/kW) |
|---|---|---|---|---|
| 0.15 | 0.01 | 0.01 | 1 | 0.01 |

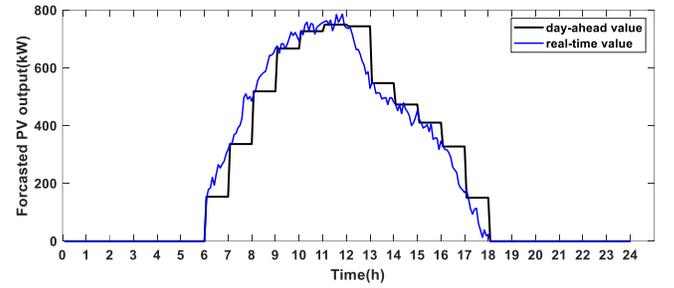

(a) forecasted day-ahead and real-time output of the solar PV unit

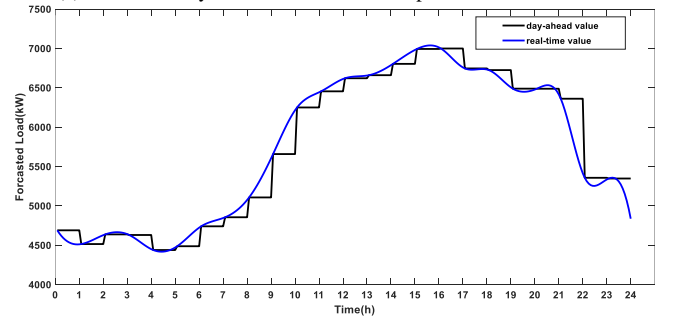

(b) forecasted day-ahead and real-time output of load

**Fig. 2**. Solar PV and load profiles

Table 4 shows the simulation results in grid-connected mode with different load uncertainty rates $\beta$ while the solar PV uncertainty rate is fixed as $\alpha = 0.05$. In this test, the solar PV uncertainty rate does not change thus the simulation results only show how the uncertainty rate of the load affects the optimal commitment and dispatch. Based on the proposed model, the BESS unit is to track the changing power of the solar PV unit while the thermal units are used to track the changing load. Therefore, the change of load uncertainty rate mostly influences the performance of the thermal units. Table 4 shows total cost of thermal units in three difference cases including the day-ahead base case, the day-ahead worst case and the sub-hourly case. As can be seen in Table 4, with the increase in load uncertainty rate $\beta$, all three total costs of thermal units increase. Total cost of thermal units in day-ahead hourly worst case and total cost in sub-hourly case are both larger than that in day-ahead hourly base case. That happens since there is uncertainty in the day-ahead hourly worst case and the forecasted values of the load in sub-hourly case are not the same as those in day-ahead base case and the thermal units need to provide extra power to balance the load. It should be noted that due to regulatory reasons, thermal units may not necessarily participate in optimizing the microgrid operation for the purpose of cost minimization. However, to show the potential interaction between the microgrid generation resources and the utility grid in satisfying microgrid loads, we have intentionally set the price of the exchange power with the grid higher than the price of the power generated by the gas unit but lower than the price of the power generated by the diesel unit.

**Table 4**
Simulation results with different load uncertainty rates $\beta$ in grid-connected mode ($\alpha = 0.05$)

| $\beta$ | Total cost of thermal units in | | |
|---|---|---|---|
| | day-ahead base case | day-ahead worst case | sub-hourly case |
| 0.05 | 11,696.67 | 16,318.41 | 12,843.41 |
| 0.10 | 12,808.84 | 17,179.10 | 13,299.01 |
| 0.15 | 13,277.92 | 18,226.73 | 13,784.34 |

**Table 5**
Simulation results with different solar PV output uncertainty rates $\alpha$ in grid-connected mode ($\beta = 0.05$)

| $\alpha$ | Total cost of BESS units in | | |
|---|---|---|---|
| | day-ahead base case | day-ahead worst case | sub-hourly case |
| 0.05 | 1.8855 | 9.486 | 3.8152 |
| 0.10 | 1.8852 | 18.972 | 3.7895 |
| 0.15 | 1.8855 | 28.458 | 3.8152 |

Table 5 shows the simulation results of the grid-connected mode with different solar PV uncertainty rates $\alpha$ and the load uncertainty rate remains unchanged at $\beta = 0.05$. In this part, the results in Table 5 show how different uncertainty rates of the solar PV unit affect the performance of the BESS unit. As mentioned above, the BESS unit is to track the changing power of the solar-storage system and thus Table 5 only shows the total cost of the BESS unit in three difference cases. Similar to the first test, the costs of the BESS unit in the day-ahead worst case and the sub-hourly case are larger than those in the day-ahead base case. The forecasted values of the solar PV unit in day-ahead worst case and in sub-hourly case are different from that in day-ahead base case, so the BESS unit should provide more power to balance the output of the solar PV unit in both day-ahead worst and sub-hourly cases. Furthermore, compared to those in Table IV, the total cost in day-ahead base case of Table 5 is just slightly different under various solar PV uncertainty rates. That means that the uncertainty of the solar PV unit has less impact on the total cost of the BESS unit in day-ahead base case than that of the loads in day-ahead base case. It is because the values of the solar PV unit output are much smaller than the values of the loads. For the same uncertainty rate, the solar PV unit causes less fluctuation for the BESS unit. Also, the solar PV unit and the BESS unit are cooperated as the solar-storage system whose output power does not have any fluctuation within one hour.

*3.2 Simulation results in islanded mode*

The proposed model applies not only in the grid-connected mode but also in the islanded mode. In this section, the simulation results in the islanded mode are presented. Table 6 shows the simulation results with different load uncertainty rates in the islanded mode. In the islanded mode, there is no power from the utility grid, so the thermal units need to produce more power to balance the load of the microgrid system. It can be observed that the total costs of the thermal units in day-head base case in Table 6 are greater than those in Table 4.

**Table 6**
Simulation results with different load uncertainty rates $\beta$ in island mode ($\alpha = 0.05$)

| $\beta$ | Total cost of thermal units in | | |
|---|---|---|---|
| | day-ahead base case | day-ahead worst case | sub-hourly case |
| 0.05 | 15,708.65 | 18,423.61 | 14,710.15 |
| 0.10 | 15,776.23 | 19,726.11 | 14,778.27 |
| 0.15 | 15,772.40 | 24,375.62 | 14,776.08 |

**Table 7**
Simulation results with different solar PV output uncertainty rates $\alpha$ in islanded mode ($\beta = 0.05$)

| $\alpha$ | Total cost of BESS units in | | |
|---|---|---|---|
| | day-ahead base case | day-ahead worst case | sub-hourly case |
| 0.05 | 4.247 | 9.486 | 4.276 |
| 0.10 | 3.466 | 18.972 | 4.089 |
| 0.15 | 3.292 | 28.458 | 4.084 |

The impact of solar PV output uncertainty as shown in Table 7 is slightly different from that of the load in the islanded mode. As indicated in constraint (60), the BESS unit tracks the output of the solar PV unit from 7 am to 6 pm in the worst case and sub-hourly case when the values of the solar PV unit output are nonzero. In addition, Eq. (59) shows that the BESS unit also participates in balancing microgrid power if needed. For the total costs of the BESS unit in day-ahead base case and sub-hourly case, the values in the islanded mode are larger than those in the grid-connected mode. That happens since there is no power imported from the grid in the islanded mode and the BESS unit must provide power to maintain the supply-demand balance. The total cost in the worst case is the same as in the grid-connected mode since the most important task for the BESS unit is to maintain the total power of the solar-storage system, which means for the same uncertainty rate, the BESS unit is capable of balancing the same amount of deviation from the solar PV output.

*3.3 Impacts of multi-timescale and regulating reserve models*

The proposed model is a multi-timescale model that considers regulating reserve explicitly. The two features including multi-timescale and regulating reserve have not been included at the same time in previously published studies. This section presents the results of four cases that use either multi-timescale model or single-timescale model and either consider regulating reserve or do not consider regulating reserve. These four cases are tested using a simulated real-time security-constrained economic dispatch (SCED) model, which is shown as follows. The sub-hourly timescale $dt$ in the tests is 5 min. A rolling time window of 25 minutes with five time-intervals ($N = 5$) is used for the real-time SCED model.

$$\min C_{real}$$
s.t. Constraints (31)-(40), (43)-(44),
$$\Delta \in \{\Delta, \Delta + 1, \dots, \Delta + N\}$$

where the objective function is:

$$C_{real} = \sum_{\Delta}^{\Delta+N} \begin{pmatrix} \sum_g C_g^P(\hat{P}_{g,t,\Delta}) \cdot \delta \\ + \sum_g |C_g^P(\hat{P}_{g,t,\Delta} - P_{g,t})| \cdot \delta \\ + \sum_b C_b^P(\hat{P}_{b,t,\Delta}^{dis} + \hat{P}_{b,t,\Delta}^c) \cdot \delta \\ + \sum_b |C_b^P(\hat{P}_{b,t,\Delta}^{dis} - P_{b,t}^{dis} + \hat{P}_{b,t,\Delta}^c - P_{b,t}^c)| \cdot \delta \\ + C_{ex}^P(\hat{P}_{ex,t,\Delta}) \cdot \delta \\ + C_{LC}^P(\widehat{LC}_{CT,t,\Delta}) \cdot \delta \end{pmatrix}$$

The first and third lines of the objective function are the costs of thermal units and BESS units in real-time dispatch, respectively. The last two lines are the cost of exchanged power in grid-connected mode and load shedding in islanded mode, respectively. To avoid big differences between the day-ahead and real-time dispatch for both thermal units and BESS units and also to reduce the impact on the grid in real-time dispatch, a penalty mechanism is included for thermal units and BESS units. It achieves the above objective functions by adding penalty costs for thermal units in the second line and for BESS units in the fourth line.

This section demonstrates how models of different timescales affects the real-time dispatch and how the consideration of regulating reserve affects the costs of the BESS units, thermal units and exchanged power in grid-connected mode or load shedding in islanded mode for real-time dispatch.

To compare the impact of multi-timescale model and single-timescale model on real-time dispatch, the two models are first run to get the unit commitment and regulating reserve, respectively. Then the corresponding unit commitment and regulating reserve are applied to real-time dispatch. Note that the real-time dispatch in the tests is based on a single-timescale (5-minute) sub-hourly model and the solar PV and load profiles in real-time dispatch are identical to those of multi-timescale model. Table 8 shows the real-time dispatch costs of the multi-timescale model and single-timescale model in both grid-connected and islanded modes. The time period chosen for the real-time dispatch tests is from hour 7 to hour 18 because solar PV unit generates non-zero power during that time period. In the grid-connected mode, there is no notable difference between the two models. The cost of the BESS unit has no difference since the BESS unit is used to track the changing output of the solar PV unit in real-time dispatch. However, the two models have big differences on the costs of thermal units and load shedding when the microgrid is operated in islanded mode. The cost of thermal units for single-timescale model is $11,279.04 while that for multi-timescale model is $8,303.76, and the difference is $2,975.28 (35.83%). There is no load shedding cost when the unit commitment and regulating reserve of multi-timescale model are applied but the load shedding cost is $1,066.91 by using single-timescale unit commitment and regulating reserve. As can be seen in this test, the results of multi-timescale model are more economical in islanded mode than those of single-timescale model. Moreover, the results of multi-timescale are also 'load-friendly' because no load shedding is needed to balance the power of the microgrid.

**Table 8**
Comparison between multi-timescale model and single-timescale model from hour 7 to hour 18 (both including regulating reserve)

| Cost terms of real-time dispatch ($) | Grid-conned Mode | | Islanded Mode | |
|---|---|---|---|---|
| | multi-timescale | single-timescale | multi-timescale | single-timescale |
| BESS unit | 10.23 | 10.23 | 10.23 | 10.23 |
| Thermal units | 6,868.78 | 6,931.22 | 8,303.76 | 11,279.04 |
| Exchanged power | 813.58 | 732.49 | 0 | 0 |
| Load shedding | 0 | 0 | 0 | 1,066.91 |

Similarly, ignoring regulating reserve causes a big difference in exchanged power with the grid in the grid-connected mode or in load shedding in the islanded mode. Table 9 shows the impact of considering regulating reserve. Note that both tests use multi-timescale model. It can be observed from Table 9 that the cost of the BESS unit, the thermal units and exchanged power in the grid-connected mode or the cost of load shedding in the islanded mode is different for the two cases, especially the cost of exchanged power in the grid-connected mode and load shedding in the islanded mode. In the grid-connected mode, the cost of exchanged power considering regulating reserve is $1,005.53 while that without considering regulating reserve is $1,275.15, which represents a 26.82% difference. In the islanded mode, the cost of load shedding considering regulating reserve is only $93.76 while that without considering regulating reserve is $14,978.60. The latter is about 160 times the former. Therefore, it is demonstrated that with regulating reserve, the results in islanded mode are more "load-friendly" due to less load shedding and lower cost.

**Table 9**
Comparison between with regulating reserve and without regulating reserve (both use multi-timescale model)

| Cost terms of real-time dispatch ($) | Grid-connected mode | | Islanded mode | |
|---|---|---|---|---|
| | With reserve | Without reserve | With reserve | Without reserve |
| BESS unit | 10.29 | 32.04 | 11.53 | 29.11 |
| Thermal units | 13,711.64 | 13,508.00 | 16,036.03 | 12,715.52 |
| Exchanged power | **1,005.53** | **1,275.15** | 0 | 0 |
| Load shedding | 0 | 0 | **93.76** | **14,978.60** |

Furthermore, Fig. 3 shows the total energy stored in the BESS unit, where the blue line is the energy in the case considering regulating reserve and the red one is that without regulating reserve. It shows that for the two models, the energies stored in the BESS unit have a large difference. The energy stored in the BESS unit considering regulating reserve is less than that without regulating reserve. That means the BESS unit considering regulating reserve is more flexible. The energy stored in the BESS unit considering regulating reserve remains unchanged from hour 9 to hour 18 in both grid-connected mode and islanded mode. During this time period, the most important role of the BESS unit is to maintain the power of the solar-storage system so the BESS unit has to prepare sufficient amount of reserve to deal with solar PV uncertainty. However, if no regulating reserve is considered, the BESS unit does not have the reserve to deal with the solar PV uncertainty. So, if the solar PV uncertainty materializes, the combined output of the solar-storage system may not remain constant. Thus, whether to consider regulating reserve or not will cause a big difference for the utility grid to decide the amount of exchanged power in the grid-connected mode or for the microgrid operator to decide the amount of load shedding in the islanded mode, and a big difference to estimate the energy stored in the BESS unit in both grid-connected and islanded modes.

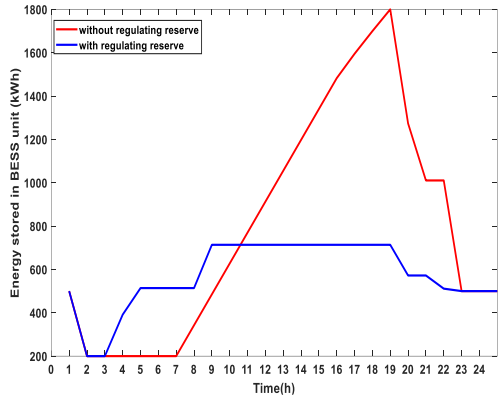
(a) Grid-connected mode

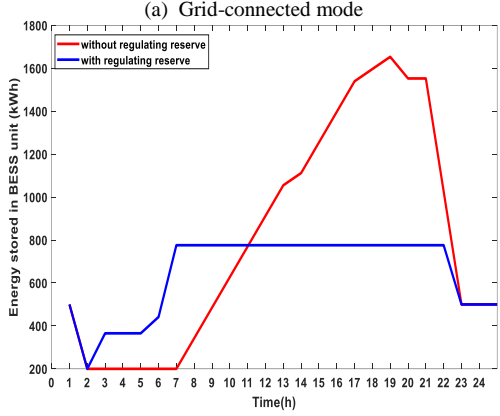
(b) Islanded mode

**Fig. 3**. Energy stored in the BESS units in different modes: (a) in grid-connected mode; (b) in islanded mode

*3.4 Simulation results for solar-storage system*

In the proposed model, the solar PV unit and the BESS unit work together as a solar-storage system. The BESS unit helps keep the combined output of the solar-storage system the same as the forecasted value of the solar PV unit. Fig. 4(a) and Fig. 4(b) show the output of the solar-storage system in the grid-connected mode and in the islanded mode, respectively. The red line is the forecasted value of the solar PV unit output and the black dotted line is the sum of the solar PV unit output and the BESS unit output in sub-hourly dispatch. It is clear to see that there is no difference between the two lines. The above simulation results show that the solar-storage system will not be affected by the operating mode.

Then, the impact of solar PV output uncertainty rate is tested in the grid-connected mode. Fig. 5(a) and Fig. 5(b) show the simulation results when $\alpha = 0.10$ and $\alpha = 0.15$, respectively. It can be observed that the two results are completely the same, and the combined solar-storage system outputs are the same as the forecasted values of solar PV unit output. The two simulation results demonstrate that the solar-storage system in this paper works well and is not affected either by operating mode or solar PV output uncertainty rate.

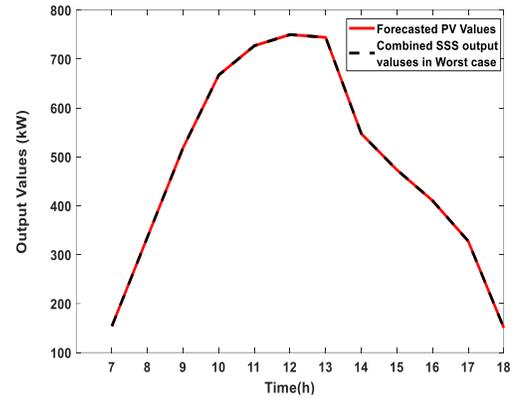
(a) in grid-connected mode

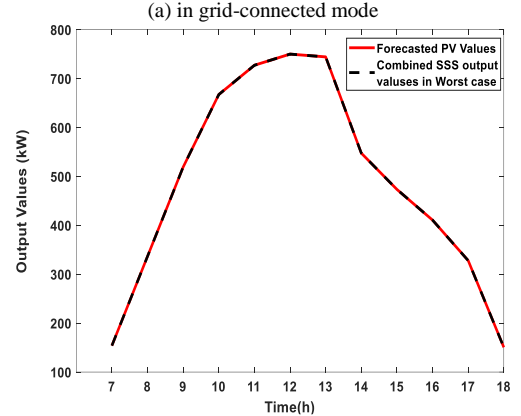
(b) in islanded mode

**Fig. 4**. Output of solar-storage system in different operating modes ($\alpha = 0.05$, $\beta = 0.15$)

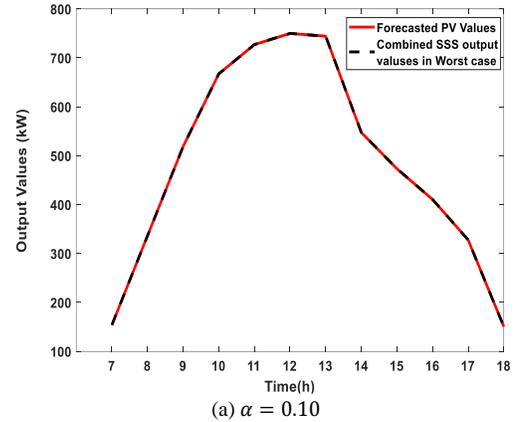
(a) $\alpha = 0.10$

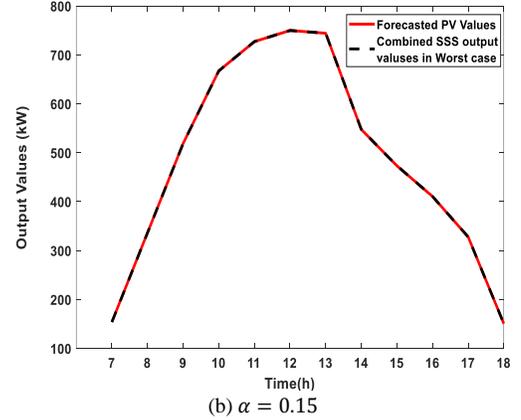
(b) $\alpha = 0.15$

**Fig. 5**. Output of solar-storage system with different uncertainty rates of PV unit

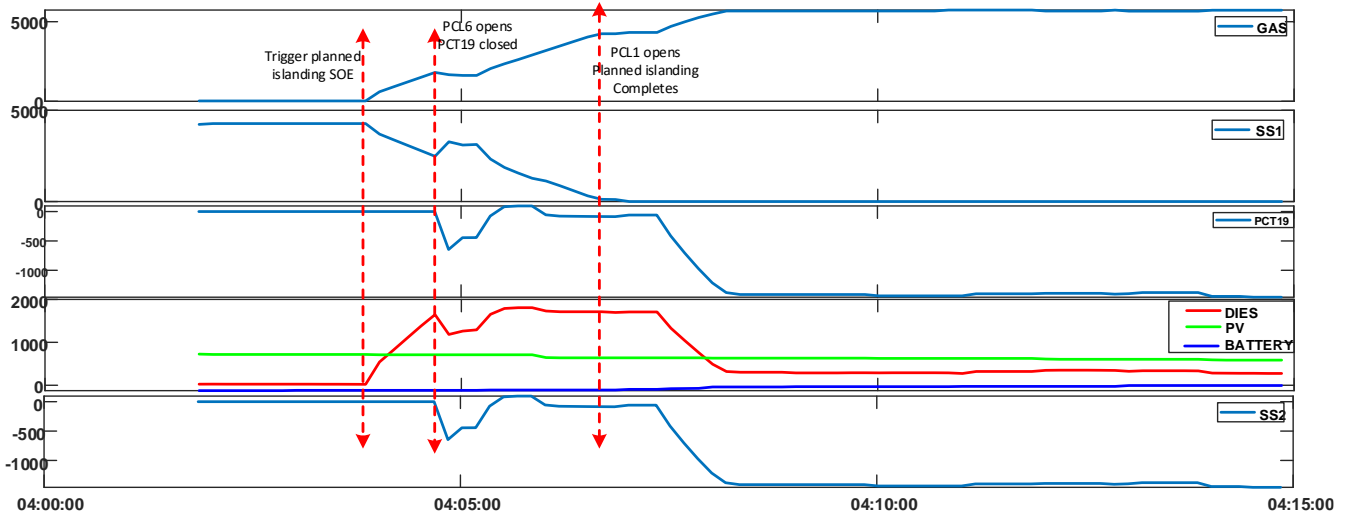

**Fig. 7**. Sample real-time dispatch test result of the solar-storage controller (planned islanding and islanded mode) (Subfigure 1: gas unit output; Subfigure 2: grid import for subsystem 1; Subfigure 3: subsystem 2 to subsystem 1 flow; Subfigure 4: diesel (orange), PV output (green), and battery (blue) output for subsystem 2; Subfigure 5: grid import for subsystem 2)

*3.5 RTDS-based testing of the proposed model*

The objective of the RTDS-based testing is to verify the proposed model in a real-world scenario. The core of the proposed model is the solar-storage coordination algorithm, which is based on robust optimization that manages the uncertainty of solar PV and load. Simply speaking, the aggregated output of the solar-storage system remains unchanged on an hourly basis, which makes solar system grid-friendly. In the testing, the targeted hourly value is assumed to be obtained based on a day-ahead hourly scheduling as discussed in the previous sections. Three tests were conducted, including (i) operation in grid-connected mode, (ii) planned islanding, and (iii) operation in islanded mode.

Fig. 6 shows a sample real-time dispatch test result of the solar-storage system in the grid-connected mode. The test runs based on the real time update of the PV from RTDS. In Fig. 6, the top red line shows the variation of the solar PV output, and the bottom blue line shows the variation of the battery output in order to maintain a pre-defined combined target (600kW in this test). Positive values of the battery indicate discharging of the battery while negative values indicate charging of the battery.

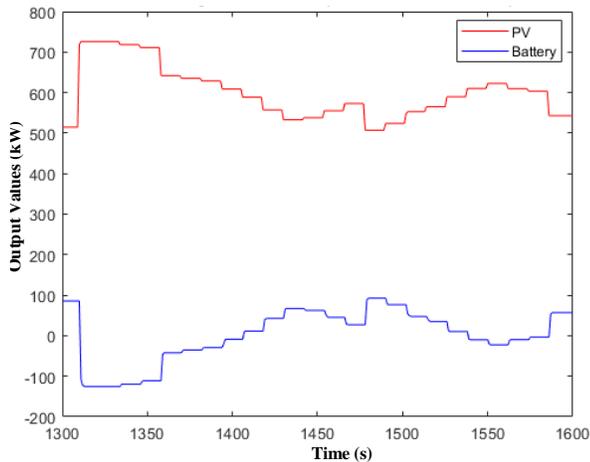

**Fig. 6.** Sample real-time dispatch test result of the solar-storage controller (grid-connected mode) (Red line: PV variation; Blue line: battery output; Target for combined PV and battery output: 600kW)

Fig. 7 shows a sample real-time dispatch test result of the solar-storage system for the planned islanding process and for the islanded mode operation. The sequence of events for the planned islanding is shown in Fig. 8. The planned islanding is initiated at 4:03:50PM. Before this, the microgrid operates at the grid-connected mode. Subsystem 1 which has a gas unit imports all its power from the utility grid to supply its load at around 4299kW. Subsystem 2 which has a diesel unit, a solar PV unit, and a BESS unit also imports all its power from the utility grid to supply its load at around 2329kW except the 600kW from the combined solar-storage system. In the grid-connected mode, the tie switch between subsystem 1 and subsystem 2 is open so there is no power exchange between the two subsystems.

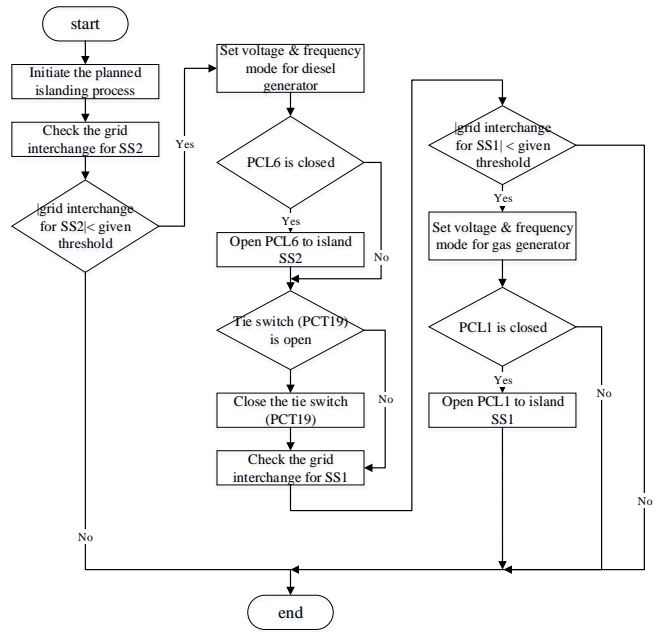

**Fig. 8**. Flow chart of Sequence of Events for Planned Islanding Process

When the microgrid master controller (MMC) receives the planned islanding signal, it conducts the redispatch required for full islanding for both subsystems, which is to balance both systems by adjusting the dispatch of the conventional generators (gas unit in subsystem 1 and diesel in subsystem 2). The setpoint for the gas unit is 4299kW to balance the load of subsystem 1. The setpoint for the diesel unit is 1729kW to balance the load of subsystem 2 along with the 600kW from the solar-storage system.

It takes about 1 minute at 4:04:50PM for the diesel unit to ramp up to balance the load in subsystem 2 and have the grid import below the threshold. Then PCL6, which is the breaker at the POI for subsystem 2, opens and the grid import to subsystem 2 becomes zero. In the meantime, PCT19, which is the tie switch connecting subsystem 1 and subsystem 2, closes and there is a flow of about 645kW from subsystem 1 to subsystem 2. The diesel unit continues to adjust its dispatch to balance the load and minimizes the flow from subsystem 1.

The gas unit in subsystem 1 continues to ramp up in order to balance the load at subsystem 1. At 4:07:00PM, the grid imports to subsystem 1 is below the threshold. Then PCL1, which is the breaker at the POI for subsystem 1, opens and the grid import to subsystem 1 becomes zero. The full islanding of the microgrid is completed.

Once the microgrid enters the islanded mode at 4:07:00PM, the MMC adjusts the output of the two conventional units (gas unit in subsystem 1 and diesel unit in subsystem 2) to minimize the cost of supplying load for the entire microgrid. As the gas unit is cheaper than the diesel unit, its setpoint is at 5700kW, which is 90% of its maximum capacity and leaves 10% as reserve. As in the grid-connected mode, the combined output of the solar-storage system remains at 600kW, the preset target value.

## 4. Conclusions

In this paper, a multi-timescale two-stage robust optimization model is proposed for the dispatch of microgrid generation resources. The proposed model combines the day-ahead hourly and sub-hourly dispatch in the first stage and evaluates the feasibility of the first-stage commitment and dispatch results under the worst-case day-ahead hourly dispatch condition caused by uncertainty of solar PV and load in the second stage. Furthermore, hourly regulating reserve is considered in the proposed model to increase the flexibility and security of the microgrid operation.

The simulation results show that the proposed model works effectively in managing the uncertainty in solar PV and load and can provide a flexible dispatch in both grid-connected and islanded modes. Due to regulating reserve, this model can deal with day-ahead dispatch with different uncertainty rates and varying real-time solar PV and load outputs. In addition, the coordinated solar-storage system can maintain the combined solar-storage output unchanged under different operating modes with different uncertainty rates. Furthermore, testing of the proposed model for the ComEd BCM at the RTDS testbed shows that the microgrid master controller equipped with the proposed model can manage a secure and economical operation for a real-life microgrid.

In the proposed model, the uncertainty for sub-hourly dispatch is not considered. Considering sub-hourly uncertainty will increase the computational complexity significantly. For future work, we will explore the potential impact of incorporating uncertainty in the real-time sub-hourly dispatch model from a practicality point of view as well as its computational implication from a solution approach perspective. We will also study the possibility of applying the proposed model on a rolling basis and implication of only considering sub-hourly constraints for the first few hours. In addition, the BESS model can be improved to consider the impact on the lifetime of the BESS. For instance, depth of discharge (DoD) is one of the most important factors that affect the lifetime of the BESS. It is negatively related to cycle time (e.g., cycle time will decrease if DoD increases). A mathematical model should be built to consider the relationship between DoD and cycle time. In that model, certain constraints about DoD and cycle time should be added to consider the impact on the lifetime of the BESS. Furthermore, RTDS is used in this paper to simulate the BCM operation. Performance analysis of the proposed model in real system operation will be conducted when BCM construction is completed.